\documentclass[proceedings]{JHEP} 
\usepackage{amsmath}
\usepackage{amssymb,amsfonts}
\usepackage{epsfig}

\newcommand\fverb{\setbox\pippobox=\hbox\bgroup\verb}
\newcommand\fverbdo{\egroup\medskip\noindent%
            \fbox{\unhbox\pippobox}\ }
\newcommand\fverbit{\egroup\item[\fbox{\unhbox\pippobox}]}
\newbox\pippobox

\newcommand{\nn}{\nonumber}

\newcommand{\C}{{\cal C}}

\newcommand{\be}{\begin{equation}}
\newcommand{\ee}{\end{equation}}
\newcommand{\ben}{\begin{enumerate}}
\newcommand{\een}{\end{enumerate}}
\renewcommand{\sp}{\ ,\qquad}
\makeatletter
\renewcommand{\@makefnmark}{\mbox{$^{\ddagger\@thefnmark}$}}
\renewcommand{\subsection}{\@startsection
  {subsection}{2}{0pt
}{-\baselineskip}{0.5\baselineskip}
  {\normalfont\normalsize\itshape}}
\makeatother
\numberwithin{table}{section}

\newcommand{\hes}{\mbox{Hes}}

\newcommand{\li}{\mathrm{Li}}
\newcommand{\old}[1]{#1_{\mathrm{old}}}
\newcommand{\oldp}[1]{(#1)_{\mathrm{old}}}
\newcommand{\gym}{g_{\mathrm{YM}}}
\newcommand{\geff}{g_{\mathrm{eff}}}

\newcommand{\aeff}{a^{\mathrm{eff}}}
\def\zz{z}    
\newcommand{\xnc}{\zz_{\mathrm{nc}}}

\title{Thermodynamics of Field Theories from Spinning Branes
\thanks{Talk presented by N.O. at {\it
Quantum aspects of gauge theories, supersymmetry and unification}, Paris,
France (Sept. 1-7, 1999).  
NBI-HE-00-11, {NORDITA-2000/15 HE}, {\tt hep-th}/0002250    
} }

\author{T. Harmark and N.A. Obers\thanks{
Work supported in part by TMR network ERBFMRXCT96-0045.}\\
    Niels Bohr Institute and Nordita, Blegdamsvej 17, DK-2100 Copenhagen, Denmark \\
    E-mail: \email{harmark@nbi.dk}, \email{obers@nordita.dk} }

\abstract{We discuss
 general spinning $p$-branes of string and M-theory
and use their thermodynamics along with the correspondence between
near-horizon brane solutions and field theories with 16
supercharges to describe the thermodynamic behavior of these
theories in the presence of voltages under the R-symmetry. The
thermodynamics is used to provide two pieces of evidence in favor
of a  smooth interpolation function between the free energy at
weak and strong coupling of the field theory. (i)  A computation
of the boundaries of stability  shows that for the
 D2, D3, D4, M2 and M5-branes
the critical values of $\Omega/T$ in the two limits are remarkably
close and (ii) The tree-level $R^4$ corrections to
the spinning D3-brane generate a decrease in the free energy at
strong coupling towards the weak coupling result.
We also comment on the generalization to spinning brane bound states
and their thermodynamics, which are relevant for non-commutative field
theories.}

\begin{document}

\maketitle 

\section{Introduction}

The discovery that
four-dimensional black holes have thermodynamic properties due to
Hawking radiation \cite{Bekenstein:1973ur}
 has been one of the main motivations to study
 the thermodynamics of black $p$-branes
 in string and M-theory \cite{Horowitz:1991cd}.
  Through the recently conjectured
 correspondence between the near-horizon limit of these brane
solutions and  quantum field theories in the large $N$ limit
\cite{Maldacena:1997re,Itzhaki:1998dd}, it has become clear that
this study not only probes the nature of  quantum gravity, but
also provides information about the thermodynamics of quantum
field theories in the large $N$ limit \cite{'tHooft:1974jz}. In particular,
for the non-dilatonic branes (D3,M2,M5)
the near-horizon limit of the supergravity solutions
 has been conjectured \cite{Maldacena:1997re}
 to be dual to a certain limit of the
 corresponding conformal field theories.
For the more general dilatonic branes of
type II string theory preserving 16 supersymmetries, similar
duality relations have also been obtained \cite{Itzhaki:1998dd}.
See Ref.  \cite{Aharony:1999ti} for a comprehensive review on the
AdS/CFT correspondence and \cite{Duff:1999rk} for lectures on branes
and AdS space.

A common feature of these dualities between near-horizon
backgrounds and field theories, is that the supergravity black
$p$-brane solution exhibits an $SO(d)$ isometry (where $d=D-p-1$
is the dimension of the transverse space) which manifests itself
as the R-symmetry of the dual field theory. As a consequence, by
considering spinning black $p$-brane solutions,  
we expect on the one hand to learn more about
the field theory side, and on the other hand, be able to perform
further non-trivial tests of the duality conjectures that include
the dependence on this R-symmetry group. The thermodynamics
on the two dual sides, including a stability analysis,
turns out to provide a useful starting point for
this comparison.
For the non-dilatonic branes this study
 has been initiated
in a number of recent papers
\cite{Gubser:1998jb,Cai:1998ji,Cvetic:1999rb}
and was further generalized in \cite{Harmark:1999xt}
for all branes that are half-BPS solutions
of string and M-theory. This work was extended in \cite{Harmark:1999rb},
 which includes the first construction of spinning brane bound states
along with their thermodynamics, which are relevant for
non-commutative field theories. In this talk, we will summarize
the main results of Refs. \cite{Harmark:1999xt,Harmark:1999rb},
focusing mainly on the spinning $p$-branes, their thermodynamics
and application to the supergravity/field theory correspondence,
while commenting at the end on the spinning brane bound states. 

\section{Spinning $p$-branes and the near-horizon limit}

The most general black $p$-brane solution
 \cite{Cvetic:1999xp,Harmark:1999rb} can be derived by oxidization
from the rotating black hole solutions in \cite{Myers:1986un}.  
We focus on general spinning black
$p$-branes that are half-BPS solutions of string and M-theory (in
the extremal and non-rotating limit), which include the M2 and
M5-branes of M-theory and the D and NS-branes of string theory.
The asymptotically-flat solution of a spinning black $p$-brane in
$D$ dimensions takes the form
\begin{eqnarray}
\label{solmet}
ds^2 &=& H^{-\frac{d-2}{D-2}}
\Big( - f dt^2 + \sum_{i=1}^p (dy^i)^2 \Big)
\nn \\ &&
+ H^{\frac{p+1}{D-2}} \Big( \bar{f}^{-1} K_d dr^2
+ \Lambda_{\alpha \beta} d\eta^\alpha d\eta^\beta \Big)
\nn \\ &&
+ H^{-\frac{d-2}{D-2}} \frac{1}{K_d L_d} \frac{r_0^{d-2}}{r^{d-2}}
\Big( \sum_{i,j=1}^n l_i l_j \mu_i^2 \mu_j^2 d\phi_i d\phi_j
\nn \\ && - 2 \cosh \alpha \sum_{i=1}^n l_i \mu_i^2 dt d\phi_i \Big)
\end{eqnarray}
where the harmonic function $H$ is given by
\begin{equation}
\label{Ldeq}
H = 1 + \frac{1}{K_d L_d} \frac{r_0^{d-2} \sinh^2 \alpha}{r^{d-2}}
\end{equation}
and
we refer to \cite{Harmark:1999xt} for further details, including
the corresponding expressions of the dilaton
$\phi$ and gauge potential $A_{p+1}$.
Since the transverse space is $d = D-p-1$ dimensional, these
spinning solutions are characterized by a set of angular momenta
 $l_i$, $i=1 \ldots n$ where $n={\rm rank} (SO(d))$,
 along with the non-extremality parameter $r_0$ and another
parameter $\alpha$ related to the charge.

Using standard methods of black hole thermodynamics we can compute
the relevant thermodynamic quantities of the general solution. The
mass $M$, charge $Q$ and angular momenta $J_i$ are determined by
the background in the asymptotic region. The Hawking-Bekenstein
entropy $S$  as well as all the intensive quantities, temperature
$T$, chemical potential $\mu$ and angular velocities $\Omega_i$
are computed from the properties at the horizon. The final results
are summarized by
\begin{eqnarray}
\label{pbranethermo}
&& M   =\rho_d  r_0^{d-2}
\Big( d-1 + (d-2)\sinh^2 \alpha \Big) \;\; 
 \\
T & = &  \frac{d-2-2\kappa }{4 \pi r_H \cosh \alpha} \ , \; \;\; 
S = 4 \pi \rho_d r_0^{d-2} r_H \cosh \alpha
\nn \\
\mu & = & \tanh \alpha \ , \; \; \;  
Q = \rho_d  r_0^{d-2} (d-2) \sinh \alpha \cosh \alpha
\nn \\
\Omega_i & =  &\frac{l_i}{(l_i^2 + r_H^2)\cosh \alpha} \ , \; \;\; 
 J_i= 2 \rho_d  r_0^{d-2} l_i \cosh \alpha \nn
\end{eqnarray}
where $r_H$ is the horizon radius determined by 
\begin{equation}
\label{horizon}
\prod_{i=1}^n \left( 1 + \frac{l_i^2}{r_H^2} \right) r_H^{d-2}=
r_0^{d-2}
\end{equation}
and $\kappa=\sum_{i=1}^n l_i^2/(l_i^2 + r_H^2) $. We have also defined
$\rho_d =V_p V(S^{d-1})/16\pi G$ with $V_p$ the
worldvolume of
the $p$-brane and $V(S^{d-1})$ the volume of the (unit radius)
transverse $(d-1)$-sphere. The quantities \eqref{pbranethermo}
obey the  conventional
Smarr formula
\begin{equation}
\label{smarr}
M = \frac{d-1}{d-2} (T S  + \Omega \cdot J) +  \mu Q 
\end{equation}
which follows from the first law of thermodynamics and the scaling behavior
of the extensive quantities.

Our main interest, however, is in the near-horizon limit  of these
spinning branes, defined as follows: One introduces
a dimensionfull parameter $\ell$ and performs the
rescaling
\begin{eqnarray}
\label{rescal} && r  = \frac{\old{r}}{\ell^2}   \sp r_0  =
\frac{\oldp{r_0}}{\ell^2}  \sp l_i = \frac{\oldp{l_i}}{\ell^2} \nn
\\ &&   h^{d-2} = \frac{\old{h}^{d-2}}{\ell^{2d-8}} \sp
   G  = \frac{\old{G}}{\ell^{2(d-2)}}
\end{eqnarray}
where  the new quantities on the left hand side are expressed in
terms of the old quantities labelled with a subscript ``old'', and
$h$ is defined by $h^{d-2} = r_0^{d-2} \cosh \alpha \sinh \alpha$.
The near-horizon limit is then defined as the limit  \( \ell
\rightarrow 0 \) keeping all the new quantities in (\ref{rescal})
fixed. The rescaling \eqref{rescal} is accompanied by appropriate
rescalings of the fields $g,\phi,A$ such that the new ones are
finite, which turns out to precisely leave the action invariant.
 The resulting near-horizon spinning $p$-brane is
  \cite{Harmark:1999xt}
\begin{eqnarray}
\label{nearmet}
ds^2 &=& H^{-\frac{d-2}{D-2}}
\Big( - f dt^2 + \sum_{i=1}^p (dy^i)^2 \Big)
 \\
 && + H^{\frac{p+1}{D-2}} \Big( \bar{f}^{-1} K_d dr^2
+ \Lambda_{\alpha \beta} d\eta^\alpha d\eta^\beta \Big)
\nn \\ &&
-2 H^{-\frac{d-2}{D-2}}
\frac{1}{K_d L_d} \frac{h^{\frac{d-2}{2}}r_0^{\frac{d-2}{2}}}{r^{d-2}}
\sum_{i=1}^n l_i \mu_i^2 dt d\phi_i \nn
\end{eqnarray}
where the harmonic function is now given by
\begin{equation}
H = \frac{1}{K_d L_d} \frac{h^{d-2}}{r^{d-2}}
\end{equation}
and we refer to \cite{Harmark:1999xt} for the corresponding expressions of the dilaton
and gauge potential.

The thermodynamics  in the near-horizon limit is then 
\begin{eqnarray}
\label{limthermo}
 & E   = \rho_d  \frac{d}{2} r_0^{d-2} & 
\\  & T = \frac{d-2-2\kappa}{4 \pi r_H}
\frac{r_0^{\frac{d-2}{2}}}{h^{\frac{d-2}{2}}} \ , \;\;\; 
S = 4 \pi \rho_d r_0^{\frac{d-2}{2}} h^{\frac{d-2}{2}} r_H & 
\nn \\ &
\Omega_i = \frac{l_i}{(l_i^2 +r_H^2)}
\frac{r_0^{\frac{d-2}{2}}}{h^{\frac{d-2}{2}}} \ , \;\;\; 
J_i = 2 \rho_d  r_0^{\frac{d-2}{2}} h^{\frac{d-2}{2}}
l_i & \nn
\end{eqnarray}
where the internal energy is obtained from the energy above
extremality $E=M-Q$. Note that since the charge and chemical
potential have become constant these are not thermodynamic
parameters anymore, so that the thermodynamic quantities are given
in terms of the $n+1$ supergravity  parameters $(r_0,l_i)$. Due to
the different scaling behavior the near-horizon quantities
\eqref{limthermo} do not satisfy the Smarr law \eqref{smarr}, but
instead the near-horizon Smarr law \cite{Harmark:1999xt}
\begin{equation}
\label{smarrnh} E = \frac{d}{2(d-2)} (TS + \Omega \cdot J)
\end{equation}
It is not difficult to obtain the energy function of the
microcanonical ensemble in terms of the extensive variables using
the horizon equation \eqref{horizon} and \eqref{limthermo}, yielding
\begin{eqnarray}
E^{d/2} & =& \left(\frac{d}{2} \right)^{d/2} \rho_d^{-(d-4)/2}
h^{-(d-2)^2/2} \left( \frac{S}{4 \pi} \right)^{d-2} \nn \\ &&
\cdot \prod_i \left( 1 + \left( \frac{2 \pi J_i}{S} \right)^2
\right)
\end{eqnarray}
Moreover, for any near-horizon spinning $p$-brane solution with
 $d$ transverse dimensions, the
Gibbs free energy takes the simple form
\begin{equation}
\label{gibbs}
 F = E - TS - \Omega \cdot J = - \rho_d  \frac{d-4}{2}
r_0^{d-2}
\end{equation}
Except for the case of one non-zero angular momentum, it is not possible in
general to obtain a closed-form expression of  $F$ in terms of the
proper intensive quantities, the temperature $T$ and the angular
velocities $\Omega_i$. However, in a low angular momentum
expansion this change of variables can be achieved for the general
spinning case to any desired order in $\omega_i=\Omega_i/T $, and
will be sufficient for some of our applications below.
 We also remark
that the Gibbs free energy \eqref{gibbs} is properly reproduced
\cite{Harmark:1999xt} for all near-horizon spinning branes by computing the
on-shell Euclidean action, taking into account the boundary term. 
This is not only an important
consistency check but also essential when calculating
string corrections to the free energy of spinning branes.

\section{Thermodynamics of dual field theories}
The thermodynamics of spinning branes and their near-horizon limit
is interesting in its own right, but also enables us to obtain
information on the thermodynamics of field theories in the
strongly coupled large $N$ limit, via the near-horizon
supegravity/field theory correspondence
\cite{Maldacena:1997re,Itzhaki:1998dd}. The fact that the branes
are spinning, introduces the new thermodynamic parameters \(
\Omega_i \) and \( J_i \) on the supergravity side which are
conjectured to correspond in the field theory to the voltage and
charge under the $SO(d)$ R-symmetry group respectively. We can trust the
supergravity description of the dual field theory, when the string
coupling \( g_s \ll 1 \) and the curvatures of the geometry are
small. This implies in all cases that the number of coincident
$p$-branes $N \gg 1$. For the M2- and M5-brane this is the only
requirement. while for the D$p$-branes  one must further demand
that \cite{Itzhaki:1998dd}
\begin{equation}
\label{geff}
 1 \ll g_{\mathrm{eff}}^2 \ll N^{\frac{4}{7-p}} \sp
 g_{\mathrm{eff}}^2 = g_{\mathrm{YM}}^2 N r^{p-3}
\end{equation}
where for the purpose of the thermodynamics, one needs to set $r=r_H$
in the effective YM coupling $\geff$.

In view of this correspondence, we can  write the Gibbs free
energy and other thermodynamic quantities in terms of field theory
variables. In particular we need to specify  the relation
between the parameter $\ell$ entering the near-horizon limit and
the relevant length scale of the theory and compute the rescaled
quantities in \eqref{rescal}. For the M2 and M5-branes the
relation to the 11-dimensional Planck length is $\ell = l_p^{3/4}$
and $l_p^{3/2}$ respectively, from which one obtains $h^6 \sim  N$
and $h^3 \sim N$. For  D-branes on the other hand, $\ell = l_s$
and one finds $G \sim g_{\rm YM}^4$ and $h^{d-2} \sim \lambda$,
where $g_{\rm YM}$ is the Yang-Mills coupling constant and
$\lambda = g_{\rm YM}^2 N$ the 't Hooft coupling.
 Restricting to
D$p$-branes, the resulting expression for the (low angular
momentum expansion of the) free energy \eqref{gibbs} in field
theory variables is then,
\begin{eqnarray}
\label{strongdpfe} &&  F_{{\rm D}p}  = - c_p V_p N^2
\lambda^{-\frac{p-3}{p-5}} T^{\frac{2(7-p)}{5-p}} \Big[ 1 +
\frac{S^1_p}{\pi^2 }
\sum_i \omega_i^2
 \nn \\ &&
 + \frac{S^2_p}{\pi^4 }
 \Big( \sum_{i} \omega_i^2 \Big)^2
+ \frac{S^3_p}{\pi^4 } \sum_i
\omega_i^4
  +   \ldots \Big]
\end{eqnarray}
where we have defined $\omega_i =\Omega_i/T$ and the numerical
coefficients $c_p$, $S_p^a$ can be found in Ref. \cite{Harmark:1999xt}. Similar
expressions for the M2 and M5 brane, proportional to $N^{3/2}$ and
$N^3$ respectively, are given in this reference as well. The same
mapping can be done for all other thermodynamic quantities
\eqref{limthermo} and we note that for the special value of $\kappa =
\frac{1}{2} (d-2)$ the temperature vanishes, implying that besides
the usual extremal limit describing zero temperature field theory,
there also exists a limit in which the temperature is zero,
accompanied by non-zero R-charges.

\section{Stability analysis}

We can now analyze the critical behavior 
\cite{Gubser:1998jb,Cai:1998ji,Cvetic:1999rb,Harmark:1999xt}  
of the near-horizon limit of
the spinning $p$-branes, which thus corresponds to the critical behavior
of the dual field theories with non-zero voltages under the R-symmetry.
The boundaries of stability have been computed in \cite{Harmark:1999xt} for both
the GCE (grand canonical ensemble with thermodynamic
variables $(T,\Omega_i)$) and the CE (canonical ensemble, with
thermodynamic variables \linebreak 
$(T,J_i)$) for $m \leq n$ equal non-zero angular
momenta.
To this end, we find the points in phase space at  which the functions
\begin{eqnarray}
\mbox{GCE:}  && \quad \det \hes (-F)  =   \frac{D_{S J }} {D_{T\Omega}}
\label{gce} \\ \mbox{CE:} && \quad
C_J = T \Big( \frac{\partial S}{\partial T} \Big)_{\{J_i\}}
  =   T \frac{D_{S J}}{D_{T J}}
\end{eqnarray}
are zero or infinite, for the GCE and CE respectively.
Here, we have  denoted by
\( D_{T \Omega} \) the determinant \( \frac{
\partial(T,\Omega_1,\Omega_2,...,\Omega_n)}
{\partial(r_H,l_1,l_2,...,l_n)} \) and likewise for \( D_{T J} \),
\( D_{S \Omega} \) and \( D_{S J} \). These determinants can be written
as finite polynomials in $l_i/r_H = J_i/S$ and their zeroes will determine
the boundaries of stability as $n$-dimensional submanifolds in the
$(n+1)$-dimensional phase diagram. Hence they crucially depend on the
fact that while there is a one-to-one correspondence between the
$n+1$ supergravity variables and the extensive quantities $(S,J_i)$,
the map to the intensive ones $(T,\Omega_i)$ or the mixed
combination $(T,J_i)$ involves a non-invertible function.

In this talk, we restrict for simplicity to the case of one non-zero
angular momentum in the GCE, in which case we find from \eqref{gce} 
that the region of stability (for $d \geq 5$) is
determined by the condition
\begin{equation}
J \leq \sqrt{\frac{d-2}{d-4}} \frac{S}{2 \pi}
\end{equation}
so that there is an upper bound on the amount of angular momentum
the brane can carry in order to be stable. Put another way, at a
critical value of the angular momentum density (which equals the
R-charge density in the dual field theory) a phase transition
occurs.
The supergravity description also determines
an upper bound on the angular velocity,
\begin{equation}
\label{ostrong}
\Omega \leq \frac{2 \pi}{ \sqrt{(d-2)(d-4)}} T
\end{equation}
which is saturated at the critical value of the angular momentum.
In Ref. \cite{Cvetic:1999rb} two scenarios have
been proposed for the nature of the resulting phase transition:
D-brane fragmentation, in which the branes fly apart in the
transverse dimension, and phase mixing in which angular momentum
localizes on the brane. The latter is possible because from
\eqref{limthermo} it follows that for each value of $\Omega/T$
between 0 and the maximum in \eqref{ostrong}, two pairs of
supergravity variables $(r_H,l)$ can be found, one corresponding
to a stable state and the other to an unstable state.

More generally, we remark that for general $d$,
the two ensembles are not equivalent and
that increasing the number of equal-valued angular momenta
enlarges the
stable region\footnote{Except for the cases $d=8,9$, which have
no stability boundary for one non-zero angular momentum.}.
We also note that the case $d=4$ is special
since the free energy \eqref{gibbs} vanishes, and it is found that
 the temperature and angular
velocity are not independent, so that the phase diagram is
degenerate; for example for one non-zero angular momentum one
finds the quadratic constraint
\begin{equation}
\label{curve} ( 2 \pi T)^2 + \Omega^2 = h^{-2}
\end{equation}
Finally, we mention that the critical exponents for all spinning branes can
be uniformly computed in both ensembles to be 1/2, a value
which satisfies scaling laws in statistical physics.

\section{Comparison to weak coupling}

An important question is to what extent do we observe the above
stability phenomena in the large $N$ limit of the dual field
theory, also at weak coupling. Extending the method
of Ref. \cite{Gubser:1998jb}, one can obtain in an ideal gas
approximation the free energies of the field theories for the
case of the M-branes and the D-branes of type II string theory,
\begin{equation}
\label{free2} F = - \tilde{c}_p  V_p T^{p+1} \sum_{\vec{\alpha}}
\li_{p+1} \left[ s_{\vec{\alpha}} \exp \Big( - \sum_{i=1}^n \alpha
\cdot \omega \Big) \right]
\end{equation}
where the $SO(d)$ R-charge weights \( \vec{\alpha} \) run over the
16 different particles and \( s_{\vec{\alpha}} \) is \( +1 \) for
bosons and \( -1 \) for fermions. The polylogarithms $\li_{p+1}$
are not defined for real numbers greater than one, but can be
continued to this region, so that the free energies can be
expressed as exact power series in $\omega_i$
\cite{Gubser:1998jb,Cvetic:1999rb,Harmark:1999xt}. 
From these one can extract the stability
behavior by considering again the zeroes of $\det( \hes (F) )$.
The resulting critical values of $\omega$ are summarized in Table
\ref{tabcompomega}, along with the corresponding values
\eqref{ostrong} at strong coupling.

\begin{table}
\begin{center}
\begin{tabular}{|c||c|c|}
\hline
Brane  & $\omega_{\mathrm{weak}}$ & $\omega_{\mathrm{strong}}$ \\ \hline
D1 & Stable   & $1.2825$ \\ \hline
D2     & $1.5404$ & $1.6223$ \\ \hline
D3     & $2.4713$ & $2.2214$ \\ \hline
D4     & $3.3131$ & $3.6276$ \\ \hline
D5 & $4.1458$ & Not defined  \\ \hline
D6     & $4.9948$ & Unstable \\ \hline
M2     & $1.5404$ & $1.2825$ \\ \hline
M5     & $4.1458$ & $3.6276$ \\ \hline
\end{tabular}
\end{center}
\caption{Comparison between the boundaries of stability
for the type II D$p$-branes n the weak
and strong coupling limits of $\lambda$
and for the M-branes
in the $N=1$ and $N\rightarrow \infty$ limits.
\label{tabcompomega}   }
\end{table}

For zero angular momentum, it has been conjectured
\cite{Gubser:1998nz} that the free energy smoothly interpolates
between the weak and strong coupling limit, which is easily extended to
the spinning case  \cite{Harmark:1999xt}. In particular, for the
D-branes  with $N$ fixed but with $\lambda$ varying between the
two limits, the conjecture reads
\begin{equation}
\label{intcon}
F_\lambda (T,\{\Omega_i\})=  f(\lambda,T,\{ \Omega_i \} )
F_{\lambda=0} (T,\{\Omega_i\})
\end{equation}
where \( F_{\lambda=0} (T,\{\Omega_i\}) \) is the free energy for
\( \lambda = 0 \). For M-branes the conjectures involves a
function interpolating between $N=1$ and any finite $N$.
An important first check of this conjecture is
the fact that the free energies of the D-branes in the two limits
show the same $N^2$ factor in front. This implies that only string
loop corrections, which carry factors of $\frac{1}{N^2}$ would
modify this behavior, and thus do not have to be considered in the
large $N$ limit.

As a corroboration of this conjecture, it is seen from Table
\ref{tabcompomega} that the critical values of the dimensionless quantity
$\Omega/T$ for the D2, D3, D4, M2 and M5-branes are remarkably
close in the weak and strong coupling limit: It is hence plausible
that in these cases the conjectured interpolation between the two
limits holds. For the D1 and D6-brane case, however, there cannot
be a smooth transition and a boundary of stability is somehow
created/destroyed at some special point between the two limits.
Finally, for the D5-brane we see that moving away from strong
coupling at some point the phase space must expand, since at weak
coupling $T$ and $\Omega_i$ are again independent thermodynamic
quantities.

Turning to the second test, we recall that the string loop
expansion in \( g_s \) and derivative expansion in \(\alpha' =
l_s^2 \) translates through the AdS/CFT correspondence   into
 a \( 1/N \) and
\( 1/\lambda \) expansion respectively. The tree-level higher
derivative term $l_s^6 R^4 $ of the type II string effective action can
thus be used to compute the $\lambda^{-3/2}$ corrections to the
free energy of the D-branes, which was analyzed for the non-rotating 
D3-brane in \cite{Gubser:1998nz}. For simplicity we focus on the
spinning D3-brane with one non-zero angular momentum, for which
the weak and strong coupling free energies are respectively given
by,
\begin{eqnarray}
\label{weakD3fe} & & F_{\lambda=0} (T,\Omega) = - N^2 V_3 T^4 \left(
\frac{\pi^2}{6} + \frac{1}{4} \omega^2 - \frac{1}{32\pi^2}
\omega^4 \right) \nn \\ \label{strongD3fe}  & & F_{\lambda=\infty}
(T,\Omega)  \nn  \\ &&  = - N^2 V_3 T^4 \left( \frac{\pi^2}{8} +
\frac{1}{8} \omega^2 + \frac{1}{16\pi^2} \omega^4   
 + \ldots  \right)
\end{eqnarray}
Since the dual theory, N=4 $D=4$ SYM, is conformal, the
interpolating conjecture \eqref{intcon} simplifies to
\begin{equation}
\label{interD3} F_\lambda(T,\Omega)  = f(\lambda,\omega)
F_{\lambda=0}(T,\Omega)
\end{equation}
From \eqref{weakD3fe}  $f(\lambda,\omega)$ is thus expected
to be smaller than one,  decreasing with $\lambda$, for fixed
$\omega < 1$, since $ -F_{\lambda=\infty} < -F_{\lambda=0} $ for
 $\omega < 1$. This has been corroborated in \cite{Harmark:1999xt} 
by computing the $R^4$ corrections to the free energy from the spinning 
D3-brane solution  \eqref{nearmet}.  

\section{Spinning brane bound states and noncommutative field theories}

We finally  comment on the generalization involving bound
states of spinning branes, which are string backgrounds with a
non-zero NSNS $B$-field (or $\C$-field in the case of M-theory).
These have recently attracted much attention, in view of the
appearance of non-commutative geometry in certain limits of such
backgrounds, as first discovered in the context of M(atrix) theory
\cite{Connes:1998cr}. More specifically, non-commutative super
Yang-Mills  (NCSYM) appears \cite{Seiberg:1999vs} in a special
low-energy limit of the world-volume theory of $N$ coinciding
D$p$-branes in the presence of a NSNS $B$-field.  This fact has
been used to extend the correspondence between near-horizon
D$p$-brane supergravity solutions and super Yang-Mills (SYM)
theories in $p+1$ dimensions
\cite{Maldacena:1997re,Itzhaki:1998dd,Aharony:1999ti},  to a
correspondence between near-horizon D$p$-brane supergravity
solutions with a non-zero NSNS $B$-field and NCSYM in $p+1$
dimensions
\cite{Maldacena:1999mh,Hashimoto:1999ut,Alishahiha:1999ci}.

By applying a sequence of T-dualities on the general spinning
D-brane solution \eqref{solmet} one obtains the general spinning D-brane
bound state solutions \cite{Harmark:1999rb}. 
The resulting backgrounds are bound states
of spinning $D(p-2k)$-branes, $k=0 \ldots m$, with $2m \leq p$ the
rank of the NSNS $B$-field,
\begin{eqnarray}
\label{solb}
 B_{2k-1,2k} &=& \tan \theta_k \Big( H^{-1} D_k - 1
\Big)
 \\
\label{ddef} D_k &=& \Big( \sin^2 \theta_k H^{-1} + \cos^2
\theta_k \Big)^{-1}
\end{eqnarray}
where $k=1 \ldots m$  and we refer to \cite{Harmark:1999rb} for the explicit
form of the complete
 solution. The extra parameters that are introduced are the angles
 $\theta_k$, $k=1 \ldots m$. Besides the charges and chemical potentials
 of the D-branes in the bound states, which depend on these angles, the
 thermodynamic quantities  of the bound state solution are
 not affected by the non-zero $B$-field and coincide with those of
 the spinning D$p$-brane given in \eqref{limthermo}.

The construction of the near-horizon limit of these solutions uses
the rescaling \eqref{rescal}, and takes the angles $\theta_k$ to
$\frac{\pi}{2}$
while keeping fixed \begin{equation} b_k = l_s^2 \tan \theta_k
\end{equation}
which are the non-commutativity parameters entering the position
commutators on the world-volume of the brane
\begin{equation}
[ y^{2k-1}, y^{2k}] \sim b_k \sp k = 1 \ldots m
\end{equation}
The dual field theories is then NCSYM in $p+1$ dimensions with
effective coupling constant
\begin{equation}
\label{geff2}
 \zz \equiv \geff^2 = \gym^2 N \left( \prod_{k=1}^m b_k \right)
r^{p-3}
\end{equation}
It turns out that the gauge coupling phase structure exhibits a far richer
structure than the commutative case, as first shown in 
\cite{Alishahiha:1999ci} and generalized in \cite{Harmark:1999rb}.   
To study this  we set $b_k = b$, $k=1\ldots m$
and consider a general path in
phase space, parametrized by 
\begin{equation}
\label{phasepath} \gym^2 \propto z^\alpha \sp r \propto z^\beta
\sp b \propto z^\gamma
\end{equation}         
keeping $N$ fixed, and where \eqref{geff2} constrains the
 scaling exponents to satisfy $\alpha + (p-3) \beta + m \gamma = 1 $.
The case $\alpha = \gamma =0$ in \eqref{phasepath} corresponds to the one
described in \cite{Alishahiha:1999ci}.  
It then follows that the supergravity description is valid provided
\cite{Harmark:1999rb} 
\begin{equation}
 \zz  \ll N^{\frac{4}{7-p}}
\left( 1 + \Big( \frac{\zz}{\xnc} \Big)^\eta \right)^{\frac{2m}{7-p}}
\end{equation}                                                           
where we have used that 
\begin{equation}
b^2 \left( \frac{r}{h} \right)^{7-p} = 
\left( \frac{\zz}{\xnc} \right)^\eta \ , \;\;  \eta = 4\beta + 2\gamma -1             
\end{equation}
Depending on the values of $\eta$ and $\xnc$ this condition can then be shown
to generate four  types of phase diagrams, and 
for each spatial worldvolume dimension of
the brane and each non-zero rank of $B$-field, a path and region
of phase space can be chosen such that the phase structure of any
of the four phase diagrams can be realized \cite{Harmark:1999rb}. 
These exhibit various interesting features,  
including regions in which the supergravity description
is valid for finite $N$ and/or  the effective coupling is allowed to
range 
from the transition point $\geff \sim 1$ all the way to infinity.  

As was argued for the
non-rotating case
\cite{Maldacena:1999mh,Alishahiha:1999ci,Barbon:1999mx,Cai:1999aw},
 the thermodynamic
quantities are the same as for commutative SYM case up to the
replacement
\begin{equation}
\gym^2 \rightarrow \gym^2 \prod_{k=1}^m b_k
\end{equation}
where $\gym$ on the left-hand side is the YM coupling constant of
the commutative theory and $\gym$ on the right-hand side of the
non-commuta{-}tive theory\footnote{For the noncommutative D3-brane case, tree-level $R^4$
corrections to the thermodynamics have been addressed in \cite{Cai:1999aw}.}. 
This was argued at weak
coupling from the field theory point of view \cite{Bigatti:1999iz}
by showing that the planar limit of SYM and NCSYM coincide. As an
application of the general phase structure analysis, the validity
of the thermodynamics for the NCSYM has been examined by requiring
that the intensive thermodynamic parameters are invariant for the
path in phase space. This condition corresponds to the
choice $\alpha + m \gamma = (5-p) \beta$ in \eqref{phasepath}.   
In particular, this enables determining the region
of phase space in which $N$ can be finite and at the same time the
coupling can be taken all the way to infinity. The resulting
condition is that \cite{Harmark:1999rb}
\begin{equation}
\label{hightemp3} \ T \gg b^{-1/2} \left(
\frac{r_0}{r_H} \right)^{\frac{7-p}{2}}
\end{equation}
showing that at fixed non-extremality parameter $ r_0$ and
horizon radius $ r_H$, the larger the non-commutativity
parameter $b$, the larger the temperature region in which these
properties are satisfied.

The presence of non-zero angular momenta does not qualitatively
change the gauge coupling phase structure, but may well provide
further insights into  NCSYM in the presence of voltages for the
R-charges. Moreover, in view of  the recent discovery that the
D6-brane theory with $B$-field decouples from gravity
\cite{Maldacena:1999mh,Alishahiha:1999ci} it is interesting 
that while the non-rotating case
 is thermodynamically unstable, for the spinning D6-brane stability
is found in the canonical ensemble for sufficiently high angular
momentum density \cite{Harmark:1999xt}. We refer to \cite{Harmark:1999rb}
 for a detailed treatment of the thermodynamics of this case.

We finally comment on some issues related to the possibility of finite $N$.
In the non-commuta{-}tive case finite $N$ is possible when 
the non-commu{-}tativity parameter $\aeff$ is large, so that
the $1/N$ expansion\footnote{For a discussion of $1/N$ corrections to
the NCSYM thermodynamics, see \cite{Barbon:1999mx} for an analysis at
strong coupling and \cite{Arcioni:1999hw} for a field theory analysis.} 
 turns into an expansion in $1/\aeff$ in that case \cite{Harmark:1999rb}. 
This suggests that for finite $N$ and large $\aeff$ planar diagrams
dominate; indeed this has recently been 
shown to hold in perturbative field theory
\cite{Minwalla:1999px}.         
The fact that finite $N$ is possible also fits together with
the recent observation \cite{Lu:1999rm} that there are  
infinitely many D$(p-2)$-branes in the near-horizon limit
of the D$(p-2)$-D$p$ system, which means  
that the world volume theory of the D$(p-2)$-brane  
has gauge group  $U(\infty)$.  
 
\section*{Acknowledgments} 
We thank the organizers of the TMR network meeting {\it
Quantum aspects of gauge theories, supersymmetry and unification} in Paris
 for a very pleasant and interesting workshop.   

\providecommand{\href}[2]{#2}\begingroup\raggedright\endgroup
\end{document}